\documentclass[reprint,amsmath,amssymb,amsfonts,superscriptaddress]{revtex4-2}
\usepackage{graphicx}
\usepackage{hyperref}
\usepackage{epigraph}
\usepackage{listings}
\newcommand{\MGMCatNLO}{MadGraph5\_aMC@NLO}
\newcommand{\pythia}{{\sc Pythia}}

\newcommand{\YPC}{Yuanpei College, Peking University, Beijing, 100871, China}
\newcommand{\DPSKLNPT}{Department of Physics and State Key Laboratory of Nuclear Physics and Technology,\\Peking University, Beijing, 100871, China}
\newcommand{\SYSU}{School of Physics, Sun Yat-Sen University, Guangzhou 510275, China}

\begin{document}
\epigraph{Time is an abstract concept created by carbon based life forms to monitor their ongoing decay.}{Thundercleese}

\title{Animating collider processes with Event-time-frame Format}
\author{Leyun Gao}
\affiliation{\YPC}\affiliation{\DPSKLNPT}
\author{Jing Peng}
\author{Zilin Dai}
\author{Sitian Qian}
\author{Tao Li}
\author{Qiang Li}
\affiliation{\DPSKLNPT}
\author{Meng Lu}
\affiliation{\SYSU}
\email{seeson@pku.edu.cn}
\email{j.peng@cern.ch}
\email{qliphy0@pku.edu.cn}
\date{\today}

\begin{abstract}
High Energy Physics processes, such as hard scattering, parton shower, and hadronization, occur at colliders around the world, e.g., the Large Hadron Collider in Europe. The various steps are also components within corresponding Monte-Carlo simulations. They are usually considered to occur in an instant and displayed in MC simulations as intricate paths hard-coded with the HepMC format. We recently developed a framework to convert HEP event records into online 3D animations, aiming for visual Monte-Carlo studies and science popularization, where the most difficult parts are about designing an event timeline and particles' movement. As a by-product, we propose here an event-time-frame format for animation data exchanging and persistence, which is potentially helpful in other visualization works. The code is maintained at \url{https://github.com/lyazj/hepani}, and the web service is available at \url{https://ppnp.pku.edu.cn/hepani/index.html}.
\end{abstract}

\maketitle

\section{Introduction}

High Energy Physics (HEP) processes occurring at colliders such as the Large Hadron Collider (LHC) are usually simulated with Markovian Monte Carlo (MC) chains, consisting of hard scattering, parton shower, and hadronization. Mature software tools are available in the market, including, {\MGMCatNLO}~\cite{2014The} and \pythia8~\cite{2015An}, to automatically simulate, e.g., central hard process and quantum chromodynamics (QCD) evolution.

In practice, various components of a physics process appear in an instant, and the intrinsic relations of particles produced are usually hard-coded within the sophisticated HepMC format output~\cite{HepMC3}, where particles are arranged according to their generation orders, and each mother-to-daughter relation is kept. Although \pythia8 provides a tool~\footnote{\url{https://pythia.org/latest-manual/examples/main300.html}} to show the event display in the form of a plane plot, it is hard to gain enough intuition due to the complicated topology and copious final state particles. In various applications after the parton shower step, rich information hidden in the Markovian history is just ignored.

In this paper, we will discuss a framework for online 3D animations based on HEP event records, aiming for visual Monte-Carlo studies and science popularization, where the most difficult parts are about designing an event timeline and particles' movement. As a by-product, we propose here an event-time-frame format for animation data exchanging and persistence, which is potentially helpful in other visualization works.

\section{From Event to Keyframes}

Conventionally in a HEP event, particle $P$ is a \textbf{mother} of $Q$, and $Q$ is a \textbf{daughter} of $P$ if particle $P$ gives birth to $Q$ and simultaneously dies. Logically, $P$ must appear earlier than $Q$ and not belong to any later generation of $Q$. Thus, by describing the whole particle system as a graph with each particle as a vertex of the graph, the system characters a directed acyclic graph (DAG), where each arc of the graph (which denotes a uni-directional relation from a vertex to another) starts from a mother and ends with one of its daughters. We call hereafter this graph as the \textbf{particle system evolution graph}, the structure of which is usually provided by most MC event generators.

Displaying the animation of a HEP event, is then naturally a problem of traversing the graph with a particularly designed strategy. One of the most direct and easy-to-implement strategies is the layer-by-layer traversal, where each layer stands for a phase (we will clarify its denotation in our algorithm below) or keyframe in the animation timeline. One must also define properly particle-to-phase correspondence or mapping, i.e., which phase each particle gets born in and which keyframe contains the newborn particle. To achieve this, we exploit an optimized Depth-First-Search (DFS) algorithm. We will explain with more details as below, on how to generate keyframes and animations from HEP event records.

\subsection{Basics: phase, birth}

\begin{figure}[tb]
\centering
\includegraphics[width=\columnwidth]{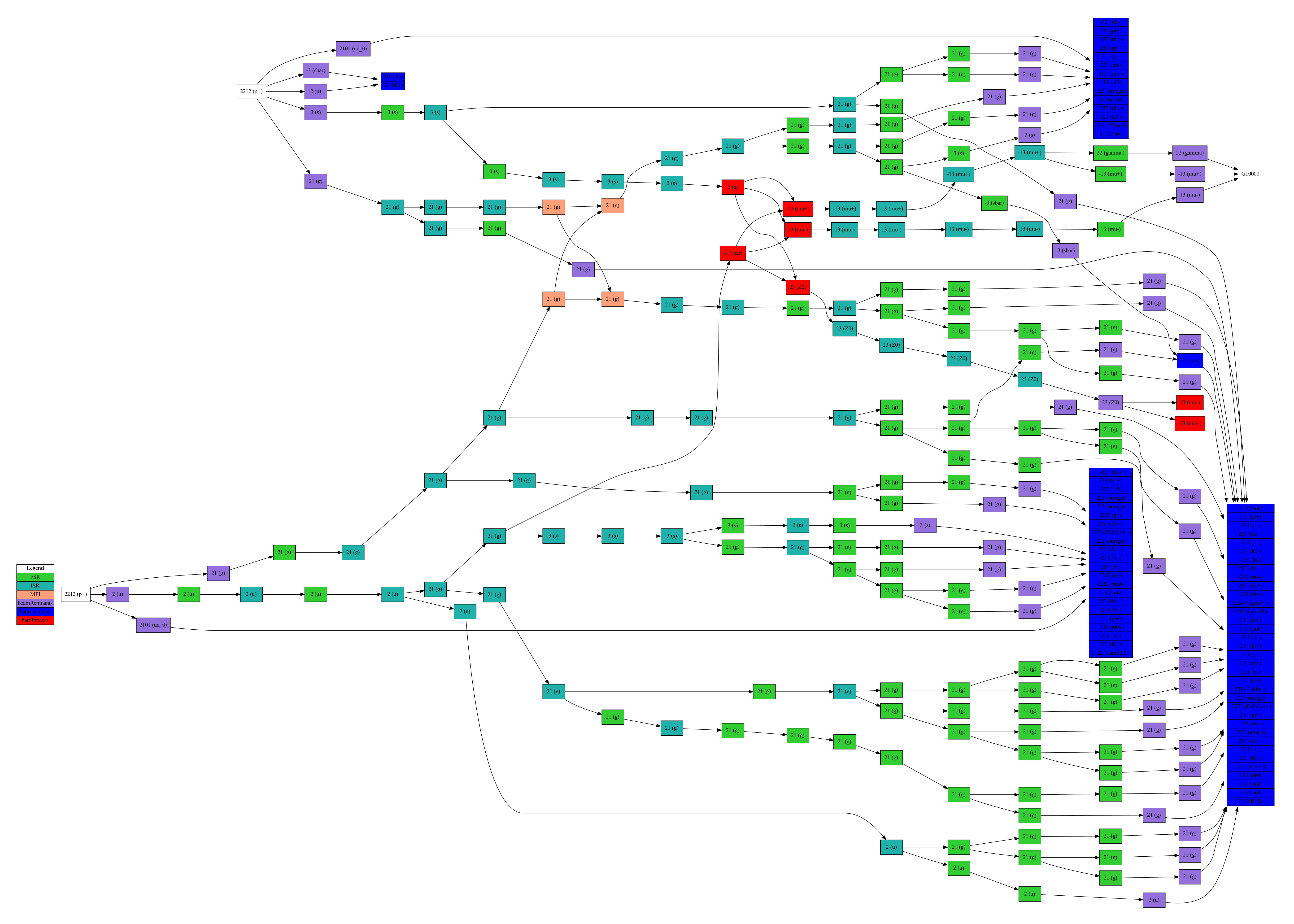}
\caption{A $\rm pp \rightarrow \mu^+\mu^-\mu^+\mu^-$ event, generated by \MGMCatNLO~with \pythia8 and visualized by \pythia8 (see \href{https://www.hepani.xyz/example/pp4mu.pdf}{pp4mu.pdf} for full size image). There are three parallel chains started from the proton in the bottom left corner. Two of them further branch, and the rightmost hadrons are one of their common terminations. There are initial state shower (ISR) products (light sea green), final state shower (FSR) products (lime green), hard process products (red) and hadronization products (blue), etc. (see \href{https://pythia.org/manuals/pythia8306/Welcome.html}{\pythia8 online manual} for details). These typical HEP processes are precisely the main orientation of our animation.}
\label{pp4mu}
\end{figure}

Each MC simulated HEP event starts from giving birth to (almost always two) original beam particles accelerated to collide. Although the beam particles have no mother particle, to make a better topological structure of the particle system evolution graph, we assign a mother named \texttt{system} numbered $0$ to them as \pythia8~\cite{PYTHIA} does. This mother is not a particle at all. However, as the only vertex with no mother and connected to all other vertices in the particle system evolution graph, \texttt{system} is the single root of the graph. From this single root, traversing the whole graph, all particles get their births while their mothers die, and the leaves (particles without any daughter) are the only final survivors. As figure \ref{pp4mu} shows, from the root to the leaves, a whole event usually branches into several parallel chains. Consider two chains that share the same start point and endpoint, chain $A\ (A_1 \rightarrow A_2 \rightarrow A_3 \rightarrow \dots \rightarrow A_m)$ and chain $B\ (B_1 \rightarrow B_2 \rightarrow \dots \rightarrow B_n)$ with $A_1 = B_1$ and $A_m = B_n$. Two ideal ``local clocks'' might indicate that the two chains started and ended simultaneously, but we would still know nothing about whether particle $A_3$ occurred earlier than particle $B_2$ based on the current MC frameworks. One of the viable visualization schemes is to let each particle occur as early as possible (see equation \eqref{birth} as an example).

To visualize the whole event, we introduce a virtual time to denote the logical order of each process and divide it into several phases. In the virtual time-space, \textbf{phase} is meant for each gap when there is no particle born, and the unsigned integer $phase$ starting from $0$ is defined as its index. Further, we regulate that a phase also includes its lower bound but excludes its upper bound, which means that there are births of particles only at the very beginnings of phases and that there's no overlap between any two phases. We then define a particle's $birth$ as the value of $phase$ at its birth. For each phase, we create a keyframe containing all newborn particles, with detailed information (i.e., mothers) about each particle included. Then while displaying animation, starting a new phase just means creating the newborn particles and removing their mothers, according to the keyframe.

Let $\mathcal M(P)$ denote the mother set of particle $P$, to make $birth(P)$ as small as possible, we define that:
\begin{equation}\label{birth}
birth(P) = \begin{cases}
0, & \mathrm{if}\ P = 0, \\
\max\limits_{M \in \mathcal M(P)}birth(M) + 1, & \mathrm{otherwise}, \\
\end{cases}
\end{equation}
where $P = 0$ means that $P$ is \texttt{system}. Equation \eqref{birth} says that in the virtual time-space, each particle gets birth later than all of its mothers, but only later than one of its mothers by one phase. Logically, no daughter can get birth before or at any of its mother's birth, so $birth(P)$ must be greater than the maximum value of its mothers' $birth$s. In equation \eqref{birth}, the difference between $birth(P)$ and the maximum value of the mothers' $birth$s is minimized to $1$, so the equation precisely gives the minimum reasonable value of $birth(P)$.

\begin{figure}[tb]
\centering
\includegraphics[width=\columnwidth]{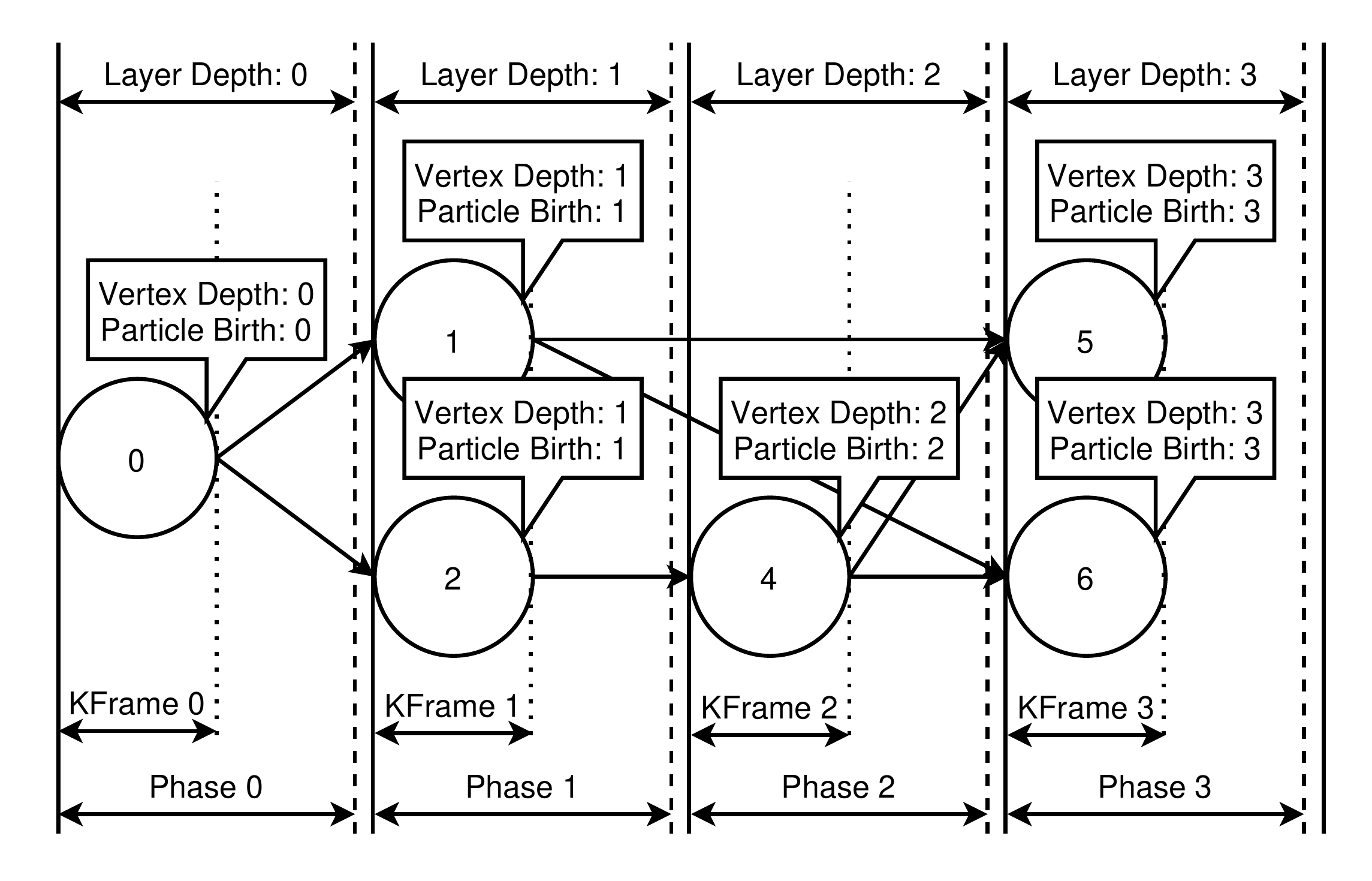}
\caption{A schematic particle system evolution graph. KFrame is for keyframe. Particles 1 and 2 belong to coming beams; particle 2 produces 4, through decay or radiation; particles 1 and 4 produce 5 and 6, through hard process or hadronization. The variable $phase$ equals the layer-by-layer traversing depth (Layer Depth in the figure), while the variable $birth$ equals the vertex depth with respect to the root $0$.}
\label{br}
\end{figure}

In this paper, for the rooted (which means the graph has a root), directed, and acyclic particle system evolution graph, we define the \textbf{depth} of a vertex as the maximum length from the root 0 to the vertex. Then by definition of the variable $birth$ from equation \eqref{birth}, a particle's $birth$ equals the corresponding vertex's depth in the particle system evolution graph. An illustration of this correspondence without loss of generality is available in figure \ref{br}.

Vertices with the same depth constitute a layer. Traversing layer-by-layer from the root 0, giving birth to each particle when visiting it, will then display the animation as one anticipates. By definition of phase, each deeper traversing layer brings a new phase into being, so the variable $phase$ must equal the current traversing depth of the particle system evolution graph. This correspondence is also illustrated in figure \ref{br}.

Direct implementation of the layer-by-layer traversal on a rooted graph is not easy. In our study, we take the keyframes as the most efficient assistance. We first calculate $birth$s for all particles, and then arrange them into keyframes according to $birth$s. With all these at hands, we are able to implement the keyframe-keyframe playing of the animation. The whole procedure is essentially equivalent to the layer-by-layer traversal of the particle system evolution graph.

Before moving to details of our implementation, we want to notice that the Breadth-First-Traversal (BFT) method is not a suitable solution to our layer-by-layer traversal problem. Take a BFT starting from the root of a rooted graph as an example, it is those vertices sharing the same minimum instead of maximum length of paths to the root that constitute a traversing layer. As shown in figure 2, for example, particles 1 and 2 will be visited in the 2\textsuperscript{nd} BFT-layer as the root 0's daughters, then particles 5, 6, and 4 will be visited in the 3\textsuperscript{rd} BFT-layer. Thus, if the designed animation is based on BFT-layer depth, particle 4 will be produced together with its daughter particles 5 and 6, which is surely not acceptable by intuition and logic.

In our case, we calculate $birth$ for all particles, by implementing equation \eqref{birth} with a recursive algorithm. Instead of relying on BFT, the implementation can be realized by reversing each arc of the graph, allowing traversing from daughters to mothers but prohibiting the opposite. The algorithm of our choice is thus equivalent to a Depth-First-Search (DFS) one, with the root $0$ as the searching target on the reversed particle system evolution graph. Notice, however, the algorithm will not terminate until all possible searching paths are traveled, in order to get the maximum length among all paths to the target. This will result in an exponential time complexity, and prevent the algorithm from being applied without proper optimization.

One of the most direct but effective optimization methods is memoization, requiring each recursion path to end with any visited vertex whose $birth$ is already evaluated. It avoids any possible duplicate evaluation of $birth$ and thus accelerates the algorithm dramatically. Applying the DFS algorithm with memoization to evaluate all $birth$s is essentially a Depth-First-Traversal (DFT) procedure on the reversed particle system evolution graph --- assuming there are $n$ particles with $m$ mother-to-daughter relations, the former procedure does the same things as the latter except for the $n$ maximum length evaluations, whose time consumption is proportional to $(n + m)$ because totally $m$ candidates are involved in comparisons to give the $n$ values. As for the DFT procedure, in our implementation, the time complexity is known to be $\mathcal O(n + m)$ because $n$ vertices are traveled through $m$ arcs, and the time consumption to get the $m$ arcs is proportional to $(n + m)$ because the $m$ arcs are accessed from the $n$ adjacency lists (adjacency list is a data structure defined in graph theory to list all adjacent vertices for a vertex). So the DFS algorithm with memoization has the overall time complexity $\mathcal O(n + m)$. It performs well enough and could be applied as the proper mapping from particles to keyframes.

\subsection{Read event data}

The particle system evolution graph could be acquired from almost all MC event generators. For example, \pythia8 is one of the most famous MC parton shower event generators~\cite{hoche2016introduction}. It can be conveniently invoked in \MGMCatNLO~ with an official plugin called mg5amc\_py8\_interface. The interface redirects logs in the standard output stream from \pythia8 into a disk file and merges the events generated into a HepMC file. In addition to HepMC records, both in version 2 and 3 from all main HEP event generators, we also enable interface with  the \pythia8 log files. However, it is worth mentioning that mg5amc\_py8\_interface overrides the status of particles to obey rules set by HepMC format. For example, many inner codes from standard \pythia8 can be overwritten as code 1, which, in HepMC records, means a particle remains stable till the end~\cite{PYTHIA}. Thus rich information can be lost, including stages (ISR/FSR/etc.) defined by \pythia8~\cite{PYTHIA}, making particles at different stages indistinguishable in the animation.

\subsection{Calculate velocity}

The velocity ($\vec v$) of a particle is derived from its relativistic momentum ($\vec p$) and relativistic energy ($e$):
\begin{equation}
\frac{\vec v}{c} = \frac{\vec p}{{\rm MeV} / c} \div \frac{e}{\rm MeV}.
\end{equation}
In some event records, energies are measured in $\rm GeV$, but momentums must be accordingly measured in $\rm GeV / c$~\cite{HepMC3}, so the equation also works in those cases. We {consider} that no momentum exchange occurs {in the lifetime of each particle,} as event generators always generate new particles to apply momentum changes~\cite{PYTHIA}. Thus particles never change their velocities in the animation.

\subsection{Find central phase and particles}

In this paper, the \textbf{central phase} denotes the phase where the hardest sub-process produces the first intermediate products (called \textbf{central particles}). As described in the \pythia8 online manual, these particles have a status whose absolute value equals 22~\cite{PYTHIA}. A sequential search algorithm is applied to find the phase, and its index is defined as $central\_phase$. Not always does an event have a (rigorous) central phase, with the status code $22$ out of presence in some cases (for example, when an intermediate particle is off-shell). However, almost all events have at least one particle with its status code in the range from $21$ to $24$, thus matching one of the four cases of the hard process particles, according to \pythia8 online manual (note that for particles of the hardest subprocess, although status codes $21$ -- $29$ are reserved, only $21$ -- $24$ have been used now)~\cite{PYTHIA}. In practice, while finding central particles, if we cannot find a particle with status code $22$, the values $23$, $21$, and $24$ are used as alternatives. If the program cannot still find a central particle, the user will get an error report.

\subsection{Build timeline and calculate positions}

To know where a phase is on the animation timeline, we need the duration of all previous phases. The variable $duration$ of a floating-point type is defined for each phase, with the default value set as 1.0 second. Users have the flexibility to change this value via a text box provided on our website. The variable $timeline$ is defined as an array-like container recording the ending of each phase. It is evaluated as:
\begin{equation}
\begin{split}
&t[-1] = 0, \\
&t[p] = t[p - 1] + duration(p), \forall p \geq 0,
\end{split}
\end{equation}
with $t = timeline$ and $p = phase$ as abbreviations. Then we can get the time span between the starting points of each of the two phases ($p_1$, $p_2$) as:
\begin{equation}
span(p_1, p_2) = t[p_1 - 1] - t[p_2 - 1].
\end{equation}
After that, the position $\vec r$ of each particle $P$ at the very beginning of each phase with $phase = p$ is calculated as:
\begin{equation}\label{r}
\vec r(P, p) = \vec r(P, birth(P)) + \vec v(P) \cdot span(p, birth(P)).
\end{equation}
Now consider $\vec r(P, birth(P))$, the position of each particle at its birth. Naturally, we expect the value to inherit from the mothers of the particle. For more than one mother, a straightforward solution is to use somehow an average value. However, in the case of hadronization, usually, a large group of mothers produces a large group of daughters. If all the daughters get birth at the mothers' mean position, they visually cluster together with the mothers distributing divergently. The phenomenon is visually unexpected and physically misleading thus should be avoided by the algorithm. In addition, we expect central particles to get birth at the center of the screen coordinates (i.e., with $\vec r = (0, 0, 0)$). Then for particles born before the central phase, their positions are natural to determine backward. If the position of a particle inherits from multiple mothers, the calculation will be hard to implement.

To solve these problems, we regulate that each particle inherits position from only one of its mothers if having more options. In such cases, the mother is called the \textbf{main mother} of the particle. As for particles with only one mother, their main mother is the only mother. Because each particle except particle $0$ has at least one mother, they all have a main mother. In order to avoid the unexpected hadron clusters on the screen, the main mother is chosen randomly and immutable since selected in case of inconsistency.

The calculation of positions starts from setting zero positions to central particles. Then we can get positions of the coming beams by several backward steps:
\begin{equation}\label{r_inherit}
\vec r(P, birth(Q)) = \vec r(Q, birth(Q)),
\end{equation}
for each particle $P$, $Q$ in particles, if $P$ is the main mother of $Q$, with $P, Q \neq 0$ and $\vec r(Q, birth(Q))$ already known. Logically, because $P \neq 0$ belongs to an ordinary particle, no $P$ will be evaluated repeatedly because of the independence of initial coming beams. Our program will check the potential logic error and avoid repeated evaluation, which will probably cause conflict results in the positions of particles. Because each particle except the particle $0$ has a main mother, and the main mother's $birth$ is less than the particle, each coming beam's position will be evaluated with equations \eqref{r} and \eqref{r_inherit} by limited steps if one of the central particles belongs to a later generation of the beam. For exception, a beam's position at its birth will remain unknown after all available evaluations if it has no association with central particles. In such unexpected cases, the user would get an error report. After all backward procedures above, we have let each particle with $birth = 1$ carry an evaluated position. Then, we do the same calculation forward sequentially:
\begin{equation}\label{r_inherit_2}
\vec r(P, birth(P)) = \vec r(Q, birth(P)),
\end{equation}
for each particle $P$, $Q$ in particles, if $Q$ is the main mother of $P$ with $birth(P) \geq 2$. Positions of some particles have been evaluated in backward procedures and will be refreshed in current forward procedure, which verifies the positions of the central particles --- if they agree with the zero values we set at first under floating-point arithmetic precision, the whole evaluation is well validated. At last, we assign
\begin{equation}
\vec r(0, 0) = (0, 0, 0).
\end{equation}
Now the position of each particle at its birth is evaluated.

Here, we have to stress again that our time and space in the animation are both \emph{virtual} and do \emph{not} explicitly have anything to do with the realistic ones. The virtual time describes the topological order of the particle system evolution, and in such time, particles move at consistent velocities, forming the virtual space. In the realistic space, particles usually do \emph{not} move with constant velocities at any (realistic) time, because of their interactions with the electromagnetic field. More importantly, almost all sub-processes of the collision are considered to occur in an instance in a narrow space, except for some long-life particles can travel a considerable distance before decaying. We have to neglect this phenomenon in realistic time-space, letting the long-life particles obey the rules of our virtual time-space, too.

\subsection{Output keyframe data}

\subsubsection{The JSON data exchange format}

JSON, a text format for structured data serialization, is now widely used for web data exchange. Compared with another web data exchange format, XML, JSON is more space-saving and  friendly to common programming languages. For example, Python3 has a built-in module {\tt json}, and JavaScript has a predefined object {\tt JSON}. They both seal convenient methods for JSON's encoding ({\tt dumps}, {\tt stringify}) and decoding ({\tt loads}, {\tt parse}). In the implementation, JSON format is used to encode the frame data of an animation.

\subsubsection{Key frame data encoding based on JSON}

A {\tt class} named {\tt Particle} is defined to store structured data of particles in keyframes. For common data exchange between programs written in different languages, a 2D array of the {\tt Particle} objects is encoded in our JSON file. The outer index of the array is the value of $birth$, while the inner one increases sequentially from zero and has no more special meaning than an index. With a unique and non-negative $birth$, each particle will be stored into the (unsigned-integer-indexed) array once and only once. When displaying animation, all the particles appear or disappear following the $birth$ or the $death$ order. The keyframes tell us how to update the particles on the screen for each new coming phase --- by adding newborn ones encoded in a keyframe and removing their mothers according to detailed information about the particles contained in the keyframe.

Notably, there's always some redundancy in a JSON file, including the one encoding our animation keyframes, as the names of each attribute of a particle repeat in each particle object (see the appendix, for example). However, after being compressed with Gzip, the size of the JSON file will be reduced by many times. If visualization is the only orientation, lower precision of floating-point numbers is sufficient. The size will be further reduced to near the Gzip-compressed HepMC record under the same compression level (see also the appendix for details).

Also, there are still other formats with better spacial efficiency, such as protobuf. For long-time storage, in addition to Gzip compression, the JSON file can be conveniently converted into protobuf or other condensed formats with appropriative libraries. However, those formats are not the best choices to encode an animation because they're usually hardly understandable to a human, not like the JSON. Generally speaking, this is also why most common Web applications now choose the JSON as their data exchange format.

In practice, a JSON file encoding an animation occupies only several tens of KiBs, approximately the size of an ordinary JavaScript file used by web pages. So we don't have to care much about the file size, but benefit more from the universality and convenience brought by the JSON format.

\section{From Keyframes to Animation}

\subsection{HTTP-server-based online animation}

An HTTP server was deployed by us with a worldwide accessible domain name \url{hepani.xyz}, powered by {\tt node.js}. A record file containing one or more MC simulated HEP events can be uploaded to the server. File type, durations, and other information will also be collected on the uploading page. The server will follow all aforementioned steps to make keyframes, use HTTP state code to indicate success or failure, and return the formatted JSON data or error message. Then, the client will be able to edit and display the animation if no error is reported.

\subsection{Particle classification}

Two approaches have been implemented by us to classify particles and give them various rendering colors and geometry sizes.

The first scheme is based on Particle Data Group (PDG) classification. As described in \emph{Monte Carlo Particle Numbering Scheme}~\cite{2000Monte}, particles of the same type are usually numbered with a consistent scheme. Particles in various modules are divided into quarks, leptons, gauge and Higgs bosons, multiple mesons, multiple baryons, etc., and the category of each particle can be directly acquired from its PDG ID.

The second scheme is based on the event process. As described by the online manual of \pythia8~\cite{PYTHIA}, a particle's status means how it was produced. For example, status $4$ or $-12$ illustrates that the particle belongs to a coming beam. Also, absolute values $21$ -- $29$ mark the hard process. To distinguish particles from different sources, they are assigned to different geometric sizes and colors. For example, particles generated from the hardest sub-process are assigned to a bigger geometric size and a red Lambert material. Same techniques are applied to the ones generated from parton showers, making our website an excellent platform to show how the parton showers occur, and what changes they make to the particle system. In the default scheme, particles considered as being produced by early sub-processes (e.g., the beam coming) are shaded more darkly, while the ones emerging in late sub-processes (e.g., particle decaying) more brightly. The scheme indicates the evolution process of the whole particle system, especially the evolution process of each Markovian chain simulating a parton shower.

As a side effect, for example, when animating two-proton collisions, the coming two protons usually decompose, radiating partons out as initial showers (ISR). At the same time, the partons change their sizes and/or colors frequently before colliding or decaying, as their status codes are modified with the particle system evolving. It may lead to the misunderstanding that two protons ``decay'' to some particles, which is not allowed in the SM. However, this problem is inevitable unless no status information is displayed. As a compromise, we provide selections to change the color and the size schemes on our website.

\subsection{Scene rendering and updating}

A famous JavaScript 3D graphics library, {\tt three.js}, is used to render particles on the website. The library {\tt three.js} implements its functions based on WebGL. The latest versions of common desktop browsers like Chrome, Edge, and Firefox have all supported WebGL, and so do many browsers on mobile platforms.

Conventionally, a mesh is required to render an object. In {\tt three.js} or WebGL, a mesh is a geometry with a material~\cite{three}. For simplification, we use the predefined sphere geometry and the Lambert material. In our implementation, the only difference between sphere geometries of different particles is the size (i.e., the argument {\tt radius}). As recommended by the official documentation of {\tt three.js}~\cite{three}, we create as few geometries and materials as possible by proper reusings of them to assemble meshes. Because of the minimal object constructions and shader compilations, the graphics performance of the website is acceptable to most desktop web browsers.

As each particle moves with a constant velocity in the lifetime, only two things are needed to convert keyframes into frames refreshed at a rate of about $60\ \rm Hz$. The first one is to update the phase and particles before each frame. A timer is set to control the animation display process. Before each frame, we will check if $phase$ needs changing (where we encounter a keyframe) as time goes on. If so, particles need to be updated and partially replaced with new ones according to the keyframe data encoded in the JSON file. The second thing is to update the position of each particle before each frame (no matter if we encounter a keyframe). We only need to calculate the time span from the latest update to the present time, multiplying it by $\vec v$ and adding it to $\vec r$ for each living particle.

By now, a complete animation is about to be created.

\subsection{More particle information on the website}

Several HTML elements are used to add an optional label to the front face of each particle (in the user's perspective). A WebGL rendered object that could make a better solution for much better performance and programmability is yet to be successfully implemented. Detailed information for each particle will appear whenever users click it. We use some HTML blocks to show basic information like $time$, $phase$, and $fps$ at the bottom of the screen. We also supply graphic time- and perspective-changing interfaces. Hotkeys are also implemented, for fast control of animation displaying, with instructions available on the help page. Now, online animation with a controllable player has been completely created.

\subsection{Animation persistence}

By now, two ways are available to make an animation persistent. First, the JSON file returned from the HTTP server can be saved directly, with keyframes included to regenerate the animation. Second, supported by the JavaScript library {\tt gif.js}, we are able to encode the animation into a GIF image file, not depending on our website and supported by most devices. The second method is under development and will soon be more practical.

\section{Examples}

\begin{figure}[tb]
\centering
\includegraphics[width=7.5cm]{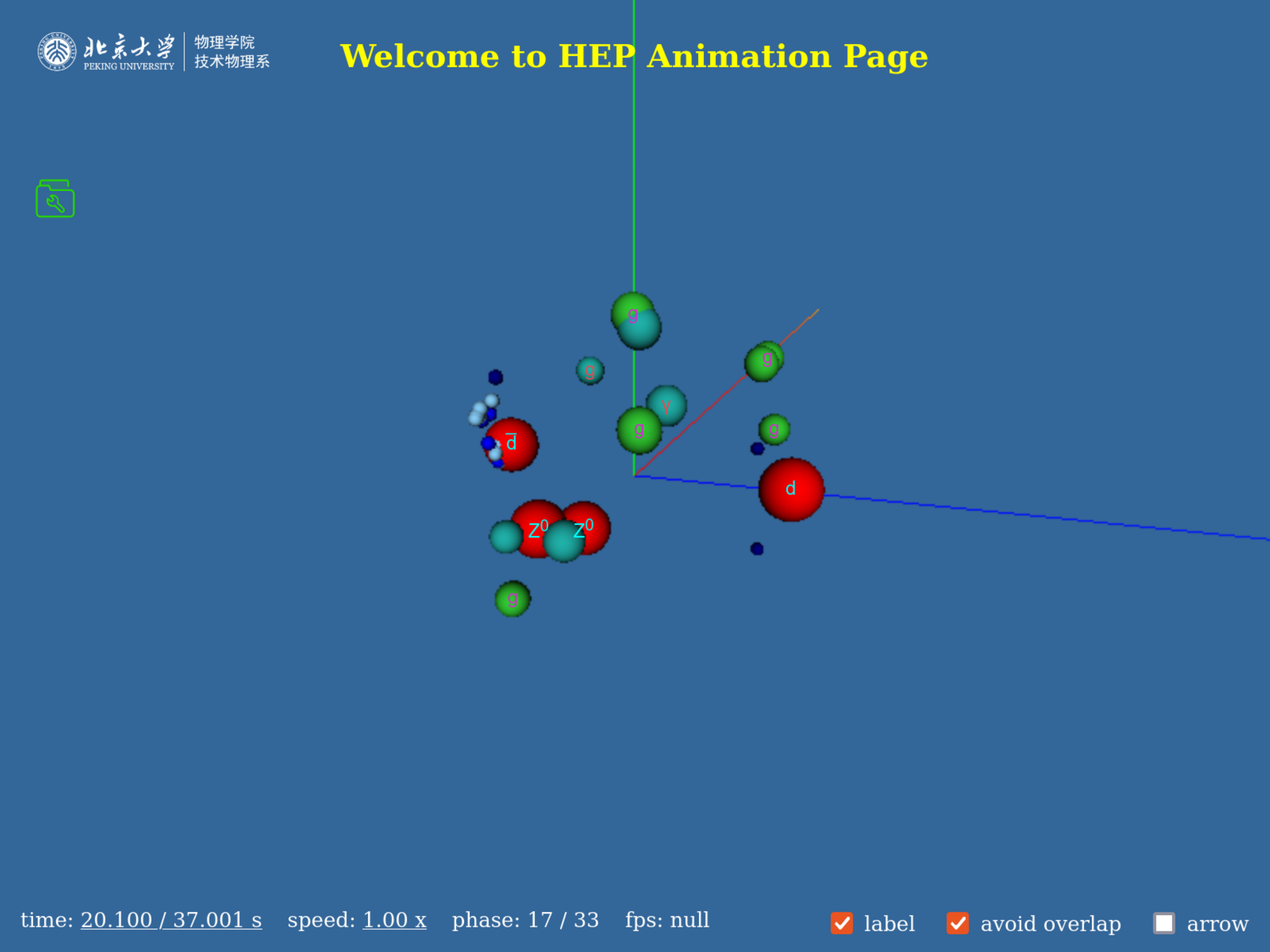}
\caption{The VBS process: phase 17 of the animation. Particles produced in the hard process are red shaded and bigger than others.}
\label{vbs_eft}
\end{figure}

\begin{figure}[tb]
\centering
\includegraphics[width=7.5cm]{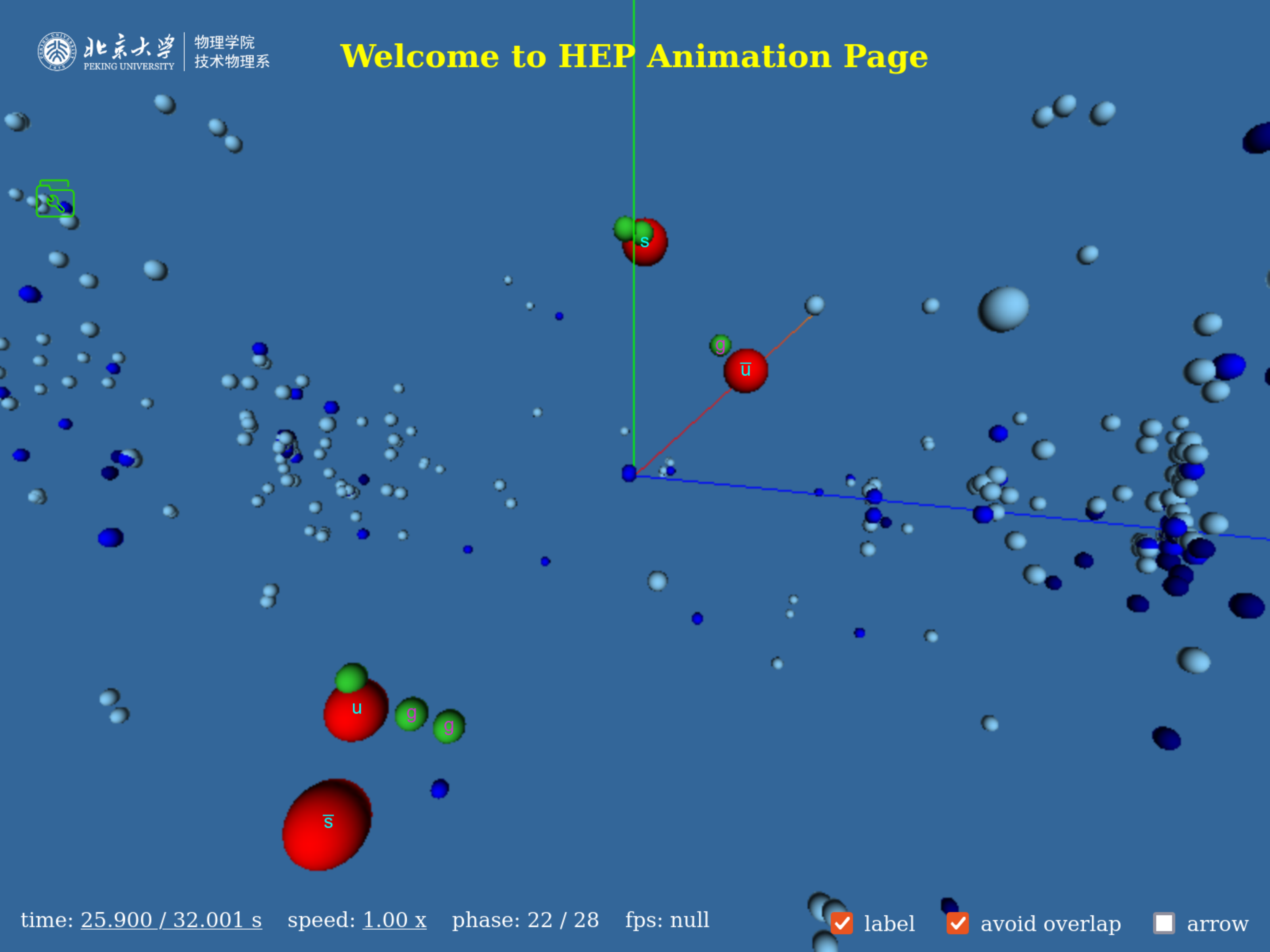}
\caption{A $\rm pp \rightarrow W^+W^-$ event: with four quarks as the final state of hard scattering and many hadrons produced by hadronization. (Dark Blue: the hadronization productions, Light Blue: the decay productions, Green: the FSR productions, Red: the hard process productions)}
\label{ppww_eft}
\end{figure}

A vector-boson scattering (VBS) process~\cite{CMS:2020fqz} for di-Z boson productions and decays into muons at the LHC is used as an example to test the website and show how 3D animations work better than plain figures for HEP popularization. Figure 3 visualizes phase 17 of the animation, with parton shower, hadronization and other processes in addition to the hard process. See the example by simply opening the link: \url{https://ppnp.pku.edu.cn/hepani/index.html}, moving to the tool icon in the top left corner of the screen, and clicking \emph{start} button.

Our website is also capable to deal with complicated final states with hadronization happening in large quantities (taking the hadronizations into consideration increases our complexity in algorithms dramatically, e.g., introduction of the main mother and the randomness in main mother choosing, but it works well then). Take figure \ref{ppww_eft}, a $\rm pp \rightarrow W^+W^-$ event~\cite{CMS:2020mxy}, for example.

For more examples, please check \url{https://www.hepani.xyz/help.html#examples}.

\section{Summary}

In this paper, we described a way to visualize HEP events as an online animation. We defined a proper mapping from particles to keyframes and provided a method to determine evolution of each particle. We introduced a JSON-encoded event-time-frame format to store the animations. We built a website for users to upload their event records and play animations supported by our HTTP cloud server. Color and size schemes are also provided to better distinguish particles at various stages in a HEP event.

The optimized Depth-First-Search keyframe generating algorithm and the JSON encoding format are expandable to support more input formats and more valuable information contained. The online animation generator and player are user-friendly designed and have so far attracted thousands of users around the word, making a good use of the advanced network and cloud technologies to contribute to HEP popularization.

While the event data format is mostly designed for collider physics, we'd also like to point out that it can be generalized to other field of physical science. For example, such data format can be applied to cosmology in visualizing the scattering of cosmic rays and the atmosphere, as well as tracing the particle behavior in simulation of material response to energetic radiation and radiation therapy.

\section*{Acknowledgments}

This work is supported in part by NSFC under Grants No. 12075004 and No. 12061141002, by MOST under grant No. 2018YFA0403900.
We would like to extend our sincere appreciation to each person or institute who accompanied, supported, or gave precious suggestions to us.

\nocite{*}
\bibliography{Animation}

\appendix

\section{Examples of the Animation JSON File}\label{JSON}
We show below the animation JSON file generated from the VBS process visualized in figure \ref{vbs_eft}. In addition to {\tt particles} and {\tt timeline}, the encoded JavaScript object also has several fields as necessary metadata.
\begin{lstlisting}
{
  "input": "py8log",
  "event": 0,
  "stamp": "Mon, 18 Apr 2022 14:38:34 GMT",
  "timeline": [0.001, 5.001, 6.001, ..., 37.001],
  "central_phase": 13,
  "central_particles": [
    {
      "no": 5,
      "id": 23,
      "name": "Z<SUP>0</SUP>",
      "status": -22,
      "colours": [0, 0],
      "r": [0.000e+00, 0.000e+00, 8.882e-16],
      "v": [7.690e-02, -4.641e-01, -4.118e-01],
      "e": 1.166e+02,
      "m": 9.098e+01,
      "birth": 13,
      "death": 14,
      "momset": [3, 4],
      "dauset": [11],
      "mainMother": 3
    },
    {
      "no": 6,
      "id": 23,
      "name": "Z<SUP>0</SUP>",
      "status": -22,
      "colours": [0, 0],
      "r": [0.000e+00, 0.000e+00, 8.882e-16],
      "v": [-3.058e-01, -3.244e-01, -5.796e-01],
      "e": 1.340e+02,
      "m": 9.141e+01,
      "birth": 13,
      "death": 14,
      "momset": [3, 4],
      "dauset": [12],
      "mainMother": 3
    }
  ],
  "central_status": 22,
  "particles": [
    [
      {"no": 0, ...}
    ],
    [
      {"no": 1, ...},
      {"no": 2, ...}
    ],
    [
      {"no": 301, ...},
      {"no": 369, ...},
      {"no": 302, ...},
      {"no": 370, ...},
      {"no": 371, ...},
      {"no": 372, ...}
    ], ...
  ]
}
\end{lstlisting}

Particles in the array \texttt{central\_particles} must have $r \sim 0$ if the particle position evaluations are properly performed. In this example, they do. Also, here the central particles both have status code $-22$, which precisely marks the hard process.

The field {\tt particles} is the mentioned 2D array of particles, with $birth$ as its outer index. Because of the limitation of space, uninteresting parts of the JSON file are omitted with ellipsis. Please check \url{https://www.hepani.xyz/example/py8log.json} for the complete version or try saving animations on our website to get more examples.

To generate JSON files with supported MC event records on your local device, please clone our github project, run {\tt make}, and execute a command like:
\begin{lstlisting}[language=bash]
"bin/Hepani" --type py8log --d0 0.001 --d1 5 \
<"example/input.txt" >"example/output.json"
\end{lstlisting}
This command works on most platforms including MS-Windows.

We'd like to end by showing TABLE \ref{tab:file-sizes} to indicate that one can benefit from the universality and convenience of the animation JSON format without concerning the file size. For each process shown in the table, we use \MGMCatNLO~3.3.2 and \pythia8 to generate a single-event HepMC record. Then we compress each of them by manually invoking \verb|gzip| 1.10 under the default compression level 6. The fact is that the derived JSON file may be slightly larger than a tiny HepMC record (like $\mathrm e^+\mathrm e^- \rightarrow \mu^+ \mu^-$, in contrary to all the other cases), but the two sizes always share the same order of magnitude.

\vspace{-2em}
\begin{table}[b]
\caption{Typical sizes of the animation JSON file (say \texttt{record.hepmc.gz}), compared with the corresponding ones of the HepMC record (say \texttt{record.json.gz}), both with level-6 Gzip compression. Each victor with a smaller value is emphasized in italic type.}
\label{tab:file-sizes}
\begin{ruledtabular}
\begin{tabular}{lrr}
event process & \texttt{record.hepmc.gz} & \texttt{record.json.gz} \\
\hline
$\mathrm{pp} \rightarrow \mathrm{ZZ}\rightarrow \mu^+\mu^-$ & 80, 534 Bytes & \emph{57, 531 Bytes} \\
$\mathrm{pp} \rightarrow \mathrm W^+\mathrm W^-$ & 49, 882 Bytes & \emph{35, 328 Bytes} \\
$\mathrm{pp} \rightarrow \mu^+\mu^-$ & 11, 753 Bytes & \emph{8, 229 Bytes} \\
$\mathrm e^+\mathrm e^- \rightarrow \mu^+\mu^-$ & \emph{656 Bytes} & 707 Bytes \\
\end{tabular}
\end{ruledtabular}
\end{table}

\end{document}